\documentclass[sigconf]{acmart}

\AtBeginDocument{%
 }

\copyrightyear{2023}
\acmYear{2023}
\setcopyright{rightsretained}
\acmConference[SIGIR '23]{Proceedings of the 46th International ACM SIGIR Conference on Research and Development in Information Retrieval}{July 23--27, 2023}{Taipei, Taiwan}
\acmBooktitle{Proceedings of the 46th International ACM SIGIR Conference on Research and Development in Information Retrieval (SIGIR '23), July 23--27, 2023, Taipei, Taiwan}\acmDOI{10.1145/3539618.3591933}
\acmISBN{978-1-4503-9408-6/23/07}

\makeatletter
\gdef\@copyrightpermission{
  \begin{minipage}{0.7\columnwidth}
   \href{https://creativecommons.org/licenses/by/4.0/}{This work is licensed under a Creative Commons Attribution International 4.0 License.}
  \end{minipage}
  \vspace{5pt}
}
\makeatother

\usepackage{caption}
\usepackage{subcaption}
\usepackage{graphicx}
\usepackage{multicol}
\usepackage{multirow}
\usepackage{adjustbox}
\usepackage{amsmath}
\usepackage{subcaption}

\usepackage{bbm}
\usepackage{adjustbox}
\usepackage{array,multirow}
\usepackage{MnSymbol}
\usepackage{pifont}

\theoremstyle{definition}
\newtheorem{definition}{Definition}[section]
\newcommand{\cmark}{\ding{51}}
\newcommand{\xmark}{\ding{55}}

\begin{document}

\title[The Role of Relevance in Fair Ranking]{The Role of Relevance in Fair Ranking}

\author{Aparna Balagopalan}
\affiliation{\institution{Massachusetts Institute of Technology}
\country{USA}
}
\email{aparnab@mit.edu}

\author{Abigail Z.\ Jacobs}
\affiliation{\institution{University of Michigan}
\country{USA}}
\email{azjacobs@umich.edu}

\author{Asia J.\ Biega}
\affiliation{\institution{Max Planck Institute for Security and Privacy}
\country{Germany}}
\email{asia.biega@mpi-sp.org}

\begin{abstract}
Online platforms mediate access to opportunity: 
relevance-based rankings create and constrain options by allocating exposure to job openings and job candidates in hiring platforms, or sellers in a marketplace. In order to do so responsibly,
these socially consequential systems employ various fairness measures and interventions, many of which seek to allocate exposure based on \emph{worthiness}. 
Because these constructs are typically not directly observable, platforms must instead resort to using proxy scores such as \emph{relevance} and infer them from behavioral signals such as searcher clicks. Yet, it remains an open question whether relevance fulfills its role as %
such a worthiness score
in high-stakes fair rankings.
\looseness=-1

In this paper, 
we combine perspectives and tools from the social sciences, information retrieval, and fairness in machine learning to derive a set of desired criteria that relevance scores should satisfy in order to meaningfully guide fairness interventions. We then empirically show that not all of these criteria are met in a case study of relevance inferred from biased user click data. We assess the impact of these violations on the estimated system fairness and analyze whether existing fairness interventions may mitigate the identified issues. Our analyses and results surface the pressing need for new approaches to relevance collection and generation that are suitable for use in fair ranking. \looseness=-1 
\end{abstract}



\maketitle

\section{Introduction}

Search systems generate rankings that do the sorting and provision of candidates to hire, of jobs to apply to, of housing options and of commercial products or services. 
 \looseness=-1
As these rankings mediate access to opportunity~\cite{biega2018equity,singh2018fairness}, various fairness measures and interventions haven been proposed to prevent harmful outcomes, such as systematic penalization of marginalized groups or new entrants to online marketplaces \cite{zeide2022silicon,xian2022opportunity,cheng2018sharing,farmaki2020airbnb,beutel2019fairness,fabbri2020effect,naghiaei2022cpfair}.
Such treatments characterize rankings as a 
causal intervention on the allocation of exposure of ranked items to searchers~\cite{joachims2021recommendations,schnabel2016recommendations} and
direct mediators of fairness~\cite{biega2018equity, singh2018fairness}, 
in line with social and legal perspectives on fairness-related harms~\cite{zeide2022silicon,xian2022opportunity}. \looseness=-1

Many existing fairness definitions propose to explicitly distribute \emph{exposure} in rankings as a function of \emph{worthiness} or \emph{deservedness}, e.g., how qualified for a job a candidate is~\cite{biega2018equity, diaz2020evaluating, singh2018fairness, morik2020controlling, bower2020individually, churilov2006towards,joachims2021recommendations}. In the fair ranking literature to date, however, the concept of deservedness has remained underdefined and has instead been represented by \emph{relevance}: Fairness definitions, such as equity of attention~\cite{biega2018equity} or fairness of exposure~\cite{singh2018fairness}, mandate that item exposure allocations in a ranking be proportional to the item's relevance to the ranking query. In socially consequential search systems, relevance thus now plays a doubly important role in the distribution of the key livelihood-impacting resources -- inequitable estimation of relevance then becomes inequitably allocated exposure. Yet, the construct of relevance in information retrieval has originally been developed to guide ordering of items to create results of high \emph{utility to searchers}. Whether the score can be \emph{repurposed as a notion of worthiness in fair rankings} remains an open question that we focus on in this paper. 

We unite perspectives on measurement theory from the social sciences \cite{jacobs2021measurement}, information retrieval, and fairness in machine learning to critically examine the role of relevance in equitable information access. We begin with a conceptual analysis of relevance, examining it from historical and definitional angles, and probing whether it can reasonably guide fairness interventions as a proxy for worthiness. The analysis allows us to derive desiderata for relevance in fair rankings: a set of conditions (credibility, consistency, stability, comparability, and availability) the scores should satisfy in order to be a reliable component of exposure-based fairness measures.

We then empirically test whether the derived criteria are met in practice in a case study of relevance inferred from biased user click data. We approach these questions as questions of long-term dynamics in a complex system that combines algorithmic and human agents; as such we conduct a simulation study with both synthetic and real world datasets, similar to past efforts
\cite{bokanyi2020understanding,cooper1973simulation,ekstrand2021simurec}. 

Click-based approaches, measuring relevance via user browsing models (i.e., conditional probabilities of relevance given a model of user clicking behavior~\cite{craswell2008experimental,wang2018position}), are indeed a common way to gather relevance at scale as it is not readily available otherwise.
\looseness=-1
Models of click behavior have been shown in prior work to sometimes poorly represent searcher behavior, reflect biases, or capture only narrow conceptions of relevance. Our results furthermore show that they can violate some of the desiderata of relevance for fair ranking. In particular, click-based relevance scores violate the criteria of availability and comparability -- as the feedback is collected only for the top-ranked items~\cite{craswell2008experimental}, accurate may not be available for most items in the corpus, and the resulting scores may not allow for accurate proportional relevance comparisons between individuals and groups.
 \looseness=-1
\looseness=-1

\looseness=-1
This paper makes the following contributions: (1) We probe the role of relevance as a proxy for worthiness in fair exposure allocation, and derive a set of five desired criteria for relevance to be a valid and reliable proxy. (2) We consider click-based relevance rankings as a case-study and observe that some desired criteria may not be met in practice. (3) We explore factors such as class imbalance or fairness interventions that may mitigate the impact of desiderata being unmet in practice, and find that impact can vary from dataset to dataset. In summary, our analyses and results demonstrate that relevance needs to be critically examined as a valid and reliable target for fair exposure allocation. The results also surface the pressing need for new approaches to relevance collection and generation (e.g., inference of relevance as calibrated probabilities), while providing a set of criteria for viable solutions.

\looseness=-1

\looseness=-1

\vspace{-1mm}
\section{Fair Ranking: The Role of Relevance}
\label{sec:related_work_fairness}

The probability ranking principle~\cite{robertson1977probability} states that an optimal ranking sorts items in order of relevance--where relevance is assumed to be unbiased and consistent.
Yet, despite this `optimal' ranking of items,
the outcomes may be unfair to different groups of users~\cite{biega2018equity, diaz2020evaluating, singh2018fairness, morik2020controlling, bower2020individually, churilov2006towards,zehlike2021fairness,yang2020causal,castillo2019fairness}.
For instance, we may observe
an unfair ranking %
when top search results for a recruiter feature primarily male candidates, despite the presence of  qualified female candidates in the selection pool. 
Many papers have proposed methods to quantify and mitigate unfairness by allocating ranking exposure in proportion to ``worthiness" of allocation~\cite{biega2018equity, diaz2020evaluating, singh2018fairness, morik2020controlling, bower2020individually, churilov2006towards,zehlike2021fairness}. Yet, the concept of worthiness remains undefined and often the relevance of an item is used as a proxy~\cite{biega2018equity,singh2018fairness,raj2022measuring}. 
In this section, we examine the role of relevance in fair ranking and probe its validity as a proxy for worthiness.
\looseness=-1
\vspace{-1mm}
\vspace{-1mm}
\subsection{Goals of Fair Exposure Allocation} 

All ranking systems and operationalizations of fairness express a normative goal. In fair ranking, different interpretations of fairness thus build from differing normative theories of discrimination underlying each framework~\cite{barocas2019fairness}. 
For example, for a goal of equal opportunity, under the view that similar items should attain equal attention, one might allocate exposure in proportion to the notion of an item's attention-deservedness or ``worthiness" at ranking time. 
Alternatively, the objective of demographic parity would be to have all groups realize their potential equally well; in implementation, this could be achieved by equalizing the levels of exposure per group. 
The ideal exposure rates for items may also be externally determined based on domain knowledge or through legal regulations: This includes,
for example, the Rooney Rule~\cite{collins2007tackling}. We direct the interested reader to recent work by ~\citet{zehlike2022fairness} which delineates the normative dimensions underlying several fair ranking techniques. 
\looseness=-1

In many fairness metrics, the \emph{relevance} of a ranked item in response to a query is considered to be a proxy for its worthiness, and the attention an item receives from searchers should ideally be proportional to its relevance~\cite{raj2022measuring,biega2018equity,singh2018fairness}.

\subsection{Measuring and Mitigating (Un)Fairness}
\label{sec:fairness_expl}

\subsubsection{Measuring Fairness}
In this paper, we focus on the family of exposure-based or attention-based metrics that allocate exposure in proportion to relevance~\cite{biega2018equity, diaz2020evaluating, singh2018fairness, morik2020controlling, bower2020individually, churilov2006towards,zehlike2021fairness}. \looseness=-1
\looseness=-1

The exposure obtained by group $G_k$ under ranking P is defined as the mean exposure obtained by all items in $G_k$:
\begin{equation}
    Exposure(G_k|P)= \frac{1}{||G_k||}  \sum_{d_i \in G_k}Exposure(d_i|P)
\end{equation}

Similarly, relevance of group $G_k$ is defined as the mean inferred relevance of all items in it: 
\looseness=-1
\begin{equation}
    Relevance (G_k)= \frac{1}{||G_k||}  \sum_{d_i \in G_k}Relevance (d_i|P)
\end{equation}

Since we consider the case of a binary protected attribute in our experiments, for simplicity all the metrics have been defined using two groups (but could be extended).
Three commonly-used metrics are:
\paragraph{Demographic fairness parity} Ratio between exposure of groups, agnostic to base rates of relevance~\cite{morik2020controlling}), measuring how much the exposure allocated to different groups varies:
\begin{equation}
      \left|\frac{Exposure(G_0|P)}{Exposure(G_1|P)} \right|
\end{equation}

\paragraph{Exposure fairness:} Ratio between exposure-relevance ratios of groups measuring how much more exposure each group receives 
proportional to its relevance~\cite{morik2020controlling}: \looseness=-1
\begin{equation}
    \frac{Exposure(G_0|P)}{Relevance (G_0|P)}
    / \frac{Exposure(G_1|P)}{Relevance (G_1|P)}
\end{equation}

\paragraph{Individual fairness:} Sum of absolute difference in the exposure and relevance scores for individual items, amortized over a sequence of rankings $[1,M]$, with $N$ items \cite{biega2018equity}.
\begin{equation}
    \sum_{i=1}^{N} \lvert \sum_{j=1}^{M}Exposure(d_i|j) - \sum_{j=1}^{M}Relevance (d_i|j) \rvert
\end{equation}

To estimate exposure for different positions, 
a weight denoting the attention a ranking position obtains from searchers on average is required. We use a log-decaying attention model for top-k positions, %
similar to the discounts in NDCG~\cite{ghosh2021fair}. All relevance and exposure scores are $[0,1]$-normalized following ~\citet{biega2018equity}. While we focus on the metrics listed above in our paper, we highlight that additional exposure-based fairness metrics have been proposed in recent literature, and direct the interested reader to work by ~\citet{raj2022measuring} for a comparitive analysis.\looseness=-1

\subsubsection{Fairness Interventions}\label{sec:fairness-intervention-highlevel}
Several methods have been proposed %
to train ranking systems that maximize fairness according to the metrics described above. 
These approaches are performed during pre-processing~\cite{asudeh2019designing,sonoda2021pre} (e.g., \citet{sonoda2021pre} consider the order of the training data), as in-processing~\cite{morik2020controlling,yang2021maximizing} (with a fairness constraint while training a learning to rank system), or as post-processing~\cite{biega2018equity,singh2018fairness,geyik2019fairness,kletti2022pareto} (re-ranking items so that system fairness is high). Notably, the expected relevance scores in the above fairness formulations may not be known a priori, and in practice one would work with \emph{estimates} from some predictive model. In particular, in the systems we consider, these scores may be obtained from the ranking model itself.
\looseness=-1

\subsection{Probing Relevance as a Proxy for Worthiness}
\label{sec:probing-relevance-as-proxy}
\subsubsection{Defining Worthiness}
\label{sec:defining_worthiness}
Fairness interventions seek to produce rankings that are more equitable, i.e., rankings that are capturing something not totally captured by inferred relevance. 
Several recent papers explicitly or implicitly refer to various notions of ``merit'' or ``worthiness'' as a basis for their intervention:
\begin{itemize}
    \item ``allocation of exposure based on merit (i.e, relevance)''~\cite{yadav2020fair},
    \item  ``We define the merit of a document as a function of its relevance to the query [$\ldots$] and we state that
each document in the candidate set should get exposure proportional to its merit”~\cite{singh2019policy},
 \item ``relevance can be thought of as a proxy for worthiness in the context of a given search task''~\cite{biega2018equity}, 
 \item ``to implement fairness constraints based on merit, we need to explicitly estimate relevance to the user as a measure of merit'' \cite{morik2020controlling}.
 \end{itemize}
In this paper, we refer to \emph{worthiness} as the underlying construct that fair rankings aim to infer and operationalize within their (possibly adjusted) relevance scores. Yet, in practice, this construct eludes an easy defintion.
Zehlike et al.\ \cite{zehlike2022fairness} note that among fairness metrics, some
consider a notion of merit when measuring disparities in exposure while mostly understanding ``merit as the utility score $Y$ at face value. However, $\ldots$the understanding of merit depends on worldviews and on one’s conception of equal opportunity.'' 
\looseness=-1

Under one interpretation, the worthiness of an item, such as a job seeker's profile on an online hiring platform, can be understood as the value from allocating attention to the item. Yet, a key question remains: the value for whom? Job seekers might receive value from being allocated attention to searchers likely to hire them. Searchers, on the other hand, might receive value from being exposed to job seekers who are qualified and likely to stay with the employer long-term. 
Worthiness scores based on the value for different stakeholders might thus diverge.

Another interpretation of worthiness, rather than tying the score to the allocation value, might look at inherent deservedness of the ranked items in a given application domain. The difference between these two interpretations is akin to the difference between outcome and process fairness; see, e.g., the discussion by~\citet{barocas2019fairness} on merit vs.\ desert for allocation.

Having laid out some of the definitional nuances of worthiness, the remainder of this section will focus on our paper's key question: Is relevance an adequate and sufficient proxy for worthiness?

\subsubsection{A Definitional Perspective on Relevance} Relevance is a construct central to all search systems. However, it is important to note that a single definition might not apply to all scenarios.  While the concept of relevance has been examined in detail before as a construct that guides information seeking behaviors~\cite{borlund2003concept, kagolovsky2001new}, several valid definitions exist, including as a user-dependent and measurable construct of information need, utility or usefulness of the viewed and assessed information object, topicality, and others~\cite{borlund2003concept,saracevic2016notion}.  
\looseness=-1
Accordingly, different definitions of relevance have been instrumented by information systems over time~\cite{saracevic2016notion}. For example, in search systems, relevance may refer to the topical match between a query and a web-page~\cite{yin2016ranking} or binary ``appropriateness" scores~\cite{kekalainen2005binary}, both often crowd-sourced and averaged~\cite{barbera2020crowdsourcing}.

In an ideal system, item ranking order should correspond to the latent relevance and satisfy a users's information need~\cite{saracevic2016notion}. Needless to say, a single-dimentional notion of relevance used during system training might not satisfy all specifications and user needs.
Relevance is a convoluted relation involving a given user's information need and the properties of ranked items -- objective scores of relevance may not exist, and instead are only defined within the context of a particular stakeholder's frame of reference~\cite{saracevic2016notion}. 
Despite this complexity, relevance is often infferred using common \emph{click measurement models} that make simplifying definitional assumptions about relevance.
\looseness=-1

\subsubsection{Measuring Relevance: From Definitions to Practice} 
Constructs such as relevance cannot be operationalized directly and are instead inferred from measurements of observable properties thought to be related to them via a measurement model~\cite{jacobs2021measurement}. 
Behaviorist search systems operationalize relevance as user engagement with ranked content, implicitly assuming that the more (latently) relevant the content is, the higher the engagement. Thus a common strategy is to estimate relevance of an item to a given query using behavioral data in the form of clicks. Clicks are translated to relevance through a user browsing model. Such models typically also account for various measurement issues, such as interaction and cognitive biases of the searchers, or the physical limitations of a ranking infrastructure. %
\looseness=-1

Another approach to measurement of relevance is crowdsourcing. Here, ``ground-truth'' relevance judgments~\cite{saracevic2016notion} are collected from human annotators for a list of items given a query. 
\looseness=-1
Because of how complex the concept of relevance is, it is however hard for external annotators to estimate another searcher's relevance ~\cite{bailey2008relevance}. Moreover, the annotation process is costly and impossible to reliably scale. Indeed, relevance judgments elicited from annotators are used primarily for system evaluation, not training~\cite{voorhees2005trec}.
\looseness=-1 %

There are several assumptions that have been traditionally made in relevance measurements (summarized from  \citet{saracevic2016notion}): that relevance is binary (ranked items can be relevant or not relevant), that relevance judgments of different items are independent, that relevance has a non-dynamic nature (i.e., values do not change over time), that relevance judgments exhibit low variance across users in non-personalized systems, and others. 
\looseness=-1
In operationalizing relevance, from the theoretical understanding to the model underlying data collection, it is crucial to note that several of these assumptions---for example, the binary nature of relevance---are still prevalent.\looseness=-1

\subsubsection{Issues with Relevance as a Worthiness Score} We hypothesize that using relevance scores as a proxy for worthiness to fairly allocate exposure has thus far been a choice of convenience. We base this claim on the fact that \emph{the construct of relevance has primarily been investigated in the context of accurate ranking construction~\cite{saracevic2016notion,borlund2003concept,kagolovsky2001new,oosterhuis2022reaching}, not as a worthiness score to guide fair exposure allocations}. 

Additionally, limitations of relevance scores~\cite{cuadra1967opening} (such as unavailability for all ranking subjects or thebinary nature of elicited judgments), of their measurement process (such as misspecifications in browsing model parameters~\cite{joachims2017unbiased}), and the consequences thereof, have also primarily been studied in the context of ranking utility. 
\looseness=-1
In the context of fairness, to address the unavailability of relevance scores, \citet{kirnap2021estimation} proposed a methodology to estimate fairness metrics using incomplete relevance judgments. However, these approximations are useful for system evaluation, not for training algorithms that fairly allocate exposure. Similarly, fair ranking evaluation benchmarks rely on exhaustive annotations that are infeasible at the scale of real systems~\cite{biega2021overview}.  \looseness=-1

In addition to the \emph{incompleteness issues}, using relevance as a proxy for worthiness involves making certain implicit assumptions about their relationship: for example, that the relative ordering of items according to worthiness is consistent with the ordering by relevance. Moreover, high variance or noisiness in relevance scores may also cause fairness measurements to be less robust or more arbitrary \cite{cooper2023variance}. It is essential to elucidate and test the assumptions of how the two concepts relate to each other if relevance is to be a \emph{valid} and \emph{reliable} candidate for approximating worthiness. 

Attending to the two-sided nature of search systems helps us further understand the way that values and positionality are embedded in relevance:  
Relevance defined as utility to the searcher/consumer and relevance defined as utility to the items/providers can be in conflict in fair ranking systems \cite{morik2020controlling,singh2018fairness}. Moreover, search systems themselves may have additional goals as well, such as maximizing the time users spend on the platform. It is thus crucial to surface \emph{whose judgment of worthiness a given relevance proxy represents}.

Table \ref{tab:rel_properties_issues} summarizes several common relevance measurement approaches and explores their potential issues as proxies of worthiness.
\looseness=-1

\begin{table*}
\begin{adjustbox}{width=\linewidth}

\begin{tabular}{ccc}
\toprule
Worthiness Proxy &  Positionality of Value & Potential Issues with Measurement  \\
\midrule
Crowd-sourced labels~\cite{barbera2020crowdsourcing,kekalainen2005binary,cuadra1967opening} & Depends on labeling instruction and perspective of labelers & May be unavailable for many items, or binary valued\\
Topical match between query and item~\cite{yin2016ranking} & Depends on attributes used & May be noisy, depends on similarity metric\\
ML-based predicted relevance~\cite{morik2020controlling} & Searcher, if inferred using interaction data & Potentially noisy, may not generalize to new queries  \\
Click-based relevance~\cite{joachims2017unbiased,agarwal2019addressing} & Searcher & Binary-valued, prone to biases, may depend on size of click logs\\
\bottomrule
\end{tabular}
\end{adjustbox}
    \caption{Potential limitations with using relevance scores as proxies for worthiness. Here, the ``positionality of value" denotes the position from which ideal ``worthiness" is judged (i.e., whose value from the ranking is considered; see Section~\ref{sec:defining_worthiness} for details).  \looseness=-1 \label{tab:rel_properties_issues} %
    }
\vspace*{-5mm}
\end{table*}

\looseness=-1

\section{Case Study: Relevance From Clicks}
\begin{figure}[h!]
\centering 
\begin{subfigure}{\linewidth}
  \centering
  \includegraphics[width=\linewidth]{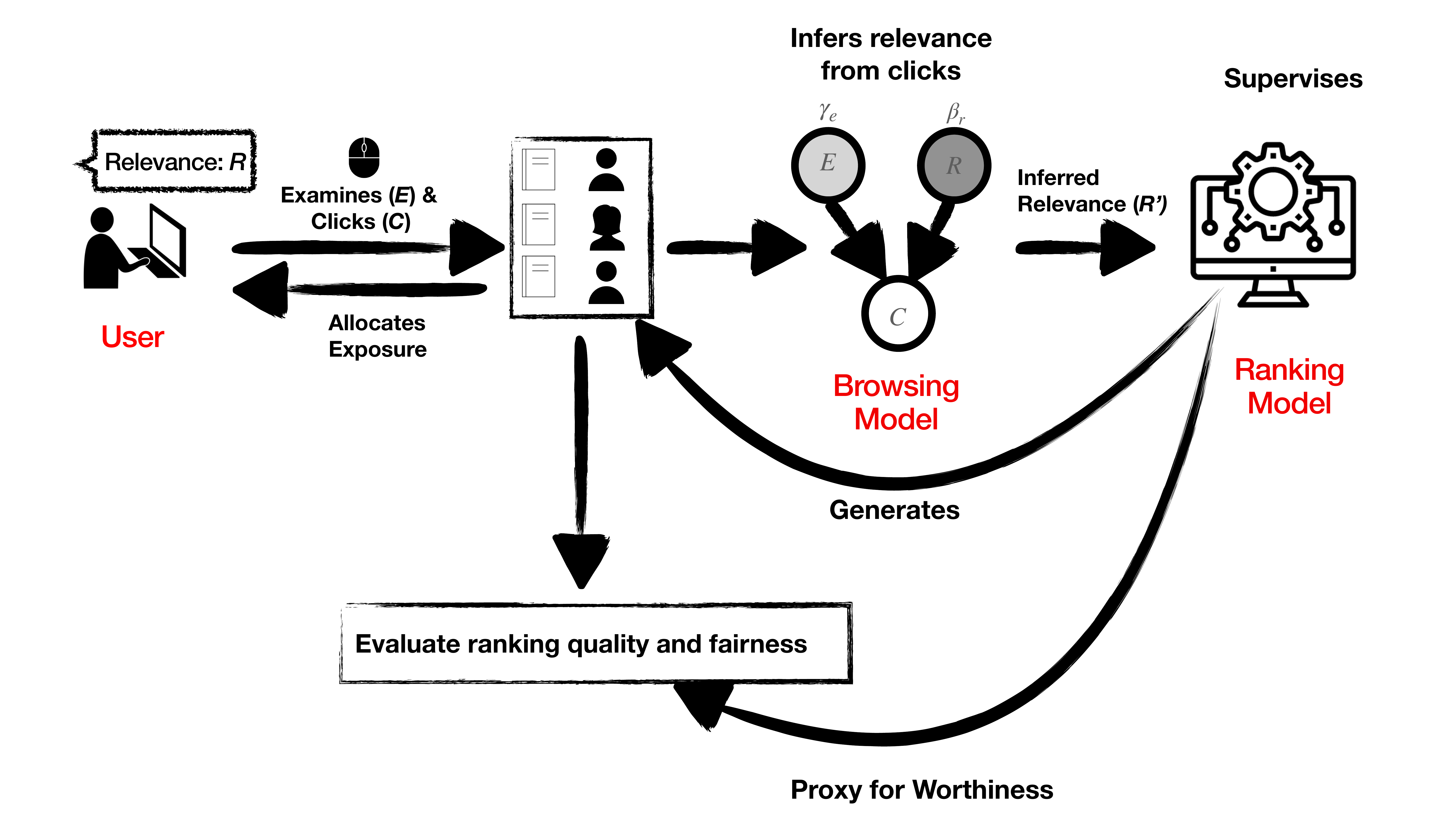}
\end{subfigure}%
 \caption{Relevance inferred using searcher clicks. A searcher inputs a query, with a latent notion of relevance ($R$) of results. Then, they examine ($E$) and click ($C$) on the ranked results, thereby allocating exposure to the items. However, they may exhibit cognitive biases (such as examining and clicking on top-ranked items with higher probability). In order to infer relevance of items from these click logs, the data has to be \emph{debiased}. Browsing models perform this operation by parameterizing the examination probability and relevance of ranked items, and then defining a model linking the click probability with these for a query and item. With such a model, the probability of an item being relevant can be estimated or relevance can be \emph{inferred}. Then, to predict the relevance for an unseen query/item, a machine learning model is trained, with loss terms influenced by parameters from the browsing model (e.g., with propensity-weighting). Finally, 
 the quality and fairness of exposure allocation is measured. \looseness=-1}
 \label{fig:relevance_from_clicks}
 \vspace*{-3.5mm}
\end{figure}

In this paper, we focus on relevance inferred from clicks as a case study.
An overview of the inference process shown in Figure~\ref{fig:relevance_from_clicks}.\footnote{Code: \url{https://github.com/Aparna-B/FairRankingRelevance}}

\subsection{Ranking %
with Click-Based Relevance}
\label{sec:ranking_from_clicks_full}
Unbiased learning-to-rank is a well studied problem in information retrieval to learn accurate rankings from biased training data
~\cite{joachims2017unbiased,wang2016learning,wang2018position,agarwal2019addressing,hu2019unbiased,vardasbi2020cascade}. Some are online ranking algorithms that involve interventions and collecting real-time user feedback~\cite{yue2009interactively,schuth2016multileave,oosterhuis2018differentiable}. Many of these systems also rely on a counterfactual model of estimations to debias the ranking loss. One common approach is to use Inverse Propensity Weighting or Score (IPS)~\cite{joachims2017unbiased} to account for various user cognitive biases in standard ranking losses, such as position bias~\cite{wang2018position}, trust bias~\cite{agarwal2019addressing}, or selection bias~\cite{melucci2016impact}. Other work proposes jointly training a ranking model and an examination propensity model~\cite{ai2018unbiased,hu2019unbiased}. \looseness=-1

Here we introduce a prototypical learning-to-rank framework
with click-based relevance, including the learning algorithms, inference methods, and evaluation metrics.

\vspace*{-2.5mm}

\subsubsection{From Clicks to Relevance via Browsing Models}
\label{sec:rel_from_clicks}
Click or browsing models -- derived using empirical studies of user information retrieval/search behavior -- make it possible to simulate
ranking systems, approximate item relevance, and evaluate search systems~\cite{chuklin2015click}. 
The relevant-intent hypothesis states that an item is clicked if and only if it is relevant and examined, which is then used to infer the latent relevance using algorithms such as Inverse Propensity Scores. 
The main variables being modelled with different assumptions in click or user browsing models is the probability of examination of items in a ranked list: $o_r$@rank $r$ is a binary variable indicating whether an item at position-$r$ is examined or not, and the relevance of items being ranked. Most browsing models parameterize the examination probability with known distributions identified using domain knowledge. When the browsing model changes, the probability of examination at a specific rank $p(o_r=1)$, and the model for user behaviour changes. As a high-level summary, a browsing model is a graphical model with two unseen nodes -- examination and relevance -- and an observable node of clicks. (For a comprehensive overview of click models, 
see~\cite{chuklin2015click}.)

In this paper, we use a position-biased browsing model of user behavior. This model assumes that examination probability is constant given a specific rank $r$. (We also note that browsing models are also used in other applications, such as debiasing user logs.)

\subsubsection{Unbiased Learning to Rank with Propensity-Weighting}
\label{sec:unbiased_ips}
The Propensity-Weighted ranking algorithm infers 
propensity weights to minimize a specific, pre-defined loss function $l(f)$ and supervise a ranking model $f$. In most ranking systems, the expected loss can be expressed as a sum of loss terms over relevant items in a list (or a pair of items). The ranking algorithm proposed by~\citet{joachims2017unbiased} uses Inverse Propensity Score (IPS) weighting to account for the biased nature of clicks (e.g., position bias as described in Section~\ref{sec:rel_from_clicks}), and weight each items's loss correspondingly. In IPS weighting, each item's loss is inversely weighted with its examination probability if it is clicked. Note that only items that are clicked contribute to this IPS-weighted loss. In the next section, we describe how the ranking model $f$ is trained and optimized. \looseness=-1

\vspace*{-1.5mm}

\subsubsection{Learning a Ranking Model}
This is the machine learning model $f$ that predicts relevance used to produce a ranking. We consider models optimized with a propensity-weighted listwise loss using the outputs of deep neural network models, following~\citet{joachims2017unbiased}. We optimize a list-wise softmax-based cross-entropy loss while training the ranking model~\cite{ai2018unbiased,ai2018learning}.
\looseness=-1
This loss is further weighted by propensity and optimized during model training, with best parameters chosen based on validation set performance. In the next section, we list evaluation metrics for assessing the utility of ranking systems.
\looseness=-1

\vspace*{-1.5mm}

\subsubsection{Performance Evaluation}
\label{sec:ndcg_expl}
Finally, the quality of the ranking algorithm is evaluated. Typically, the ranking produced by the model is compared against a ground truth ranking (assumed to be an unbiased gold standard). In this paper we focus on the Normalized Discounted Cumulative Gain (NDCG), a normalized measure of ranking quality favoring occurrence of more relevant documents in the higher ranking positions through a logarithmic discount~\cite{jarvelin2017ir}.\looseness=-1

\section{Desiderata of Relevance in Fair Ranking}
\label{sec:desiderata_list}
In this section, we describe a set of {desiderata} that the relevance scores  ought to satisfy to be meaningful and useful for fairly allocating exposure. Drawing on domain knowledge about relevance-based ranking systems and the assumptions within,
these desiderata offer a pathway towards \emph{valid} and \emph{reliable} measures of relevance, where the properties of proxy scores should match their theoretical ideal across a range of qualitative dimensions \cite{jacobs2021measurement}. \looseness=-1

In the definitions below, let $w$ denote true relevance (worthiness) scores, and $r$ denote inferred relevance scores, i.e., the measurements of $w$ that will be used to allocate exposure. $K$ denotes the number of items to be ranked. \looseness=-1

\vspace*{-2.4mm}
\paragraph{Credibility} For inferred relevance scores to be useful, they must behave as we expect. In a ranking setup, we can ask: do items with high true relevance (worthiness) have higher inferred relevance scores across runs? (And lower, lower?) 
Absence of credibility would imply that the inferred relevance scores are not capturing how the `true' relevance scores would be expected to, and so lacks face validity and content validity---and may even lack reliability across similar inputs to the model~\cite{jacobs2021measurement}. Thus, it might be an unsuitable proxy under views of fairness such as ``equal opportunity" where exposure should ideally be allocated in proportion to worthiness. \looseness=-1
\begin{definition}[Credibility]
If $i$ and $j$ are two items to be ranked, credibility necessitates, in expectation:
\begin{equation}
    w_{i} \geq w_{j}  \equiv r_{i} \geq r_{j}
\end{equation}
\end{definition}
\vspace*{-2mm}
\paragraph{Consistency} A core assumption in unbiased learning-to-rank is that of consistency: in the limit of sufficient data, the estimated expected relevance scores converge. Thus, meeting this statistical property necessitates the criterion: do the inferred relevance scores converge? If this criterion remains unmet, then fair ranking metrics relying on relevance may not converge either. \looseness=-1
\begin{definition}[Consistency]
In the limit of sufficient data, the predicted scores converge within an acceptable error range. Let $N$ denote the number of expected training steps (e.g., the size of the training set), and $\hat{r}_i$ the predicted relevance score for item $i$ at the $N$th training step. If $S_n=\frac{1}{K}\sum_{i=1}^{K}[(r_i^{n} - \hat{r}_i)^2]$ measures the mean squared deviation between relevance scores obtained by training the relevance prediction model for $n$ steps and $\{\hat{r_i}\}_{i=1}^{K}$, and $\epsilon$ is the acceptable error value (close to 0), then consistency requires:

\begin{equation}
    \exists N_0 \textnormal{ such that } S_n \leq \epsilon \forall n\geq N_0
\end{equation}
\end{definition}
\vspace*{-2mm}
\paragraph{Stability} Relevance ought to have {test-retest reliability}, i.e., that measurements from the same model for the same input do not vary more than an expected limit (e.g., due to unavoidable stochasticity). In the ranking setup, the criterion to test then is: do the inferred relevance scores vary more than an established limit across runs with minor variations in stochastic parameters or initialization? This is related to robustness and how noisy estimates of fairness would be. \looseness=-1
\begin{definition}[Stability]
If $n$ denotes the number of experimental runs with variations in unimportant parameters, and $\epsilon$ is the acceptable variation across runs, then:
\begin{equation}
    \frac{1}{K}\sum_{j=1}^{K}[\frac{1}{n}\sum_{i=1}^{n}[(r_{ij} - \mu_{j})^2]] \leq \epsilon
\end{equation}
where $\mu_j=\frac{1}{n}\sum_{i=1}^{n} r_{ij}$, and $r_{ij}$ denotes relevance for the $j$th item under the $i$th run. 
\end{definition}

\vspace*{-2mm}
\paragraph{Comparability} Several fairness interventions suggest that exposure should be allocated in proportion to inferred relevance~\cite{biega2018equity,singh2018fairness}, with the underlying assumption that this proportionality would hold for true relevance (worthiness) of items/groups. Thus we must verify that the relative ratios of relevance scores match the corresponding ratios of true worthiness, i.e., that our model has structural validity---that these relative properties behave as expected~\cite{jacobs2021measurement}.
For group fairness, the average relevance is taken as the aggregate across all items belonging to a group. For individual fairness, this criterion is evaluated at the per-item level.\looseness=-1
\vspace*{-2mm}

\begin{definition}[Comparability]
If $i$ and $j$ are two items or groups of items (e.g., defined on the basis of a sensitive attribute) to be ranked, comparability necessitates:
\begin{equation}
    \frac{w_i}{w_j} \approx t \equiv \frac{r_i}{r_j} \approx t
\end{equation}
\end{definition}

\paragraph{Availability} Past research in Information Retrieval (IR) has established that searchers exhibit position-bias in viewing and inspecting ranked results~\cite{craswell2008experimental}. As a result, this variance in attention across different positions is modeled while allocating exposure for an item or group of ranked items. Then, exposure is allocated in proportion to uncorrected (i.e., without any bias-correction) relevance scores. Thus, an underlying assumption is that relevance scores are \emph{available} for all items, and are \emph{unbiased} estimates of true relevance (or worthiness). The availability criterion tests this assumption that unbiased inferred relevance scores are available for all items---without which it would be difficult to claim validity or reliability of relevance scores.\looseness=-1

\begin{definition}[Availability]
If $f$ is a function that outputs $1$ if $f(x)$ is defined and $0$ if not, then, 
\begin{equation}
    f(r_i)=1 \forall i \in \{1,2 \dots K\}
\end{equation}
\end{definition}
Additionally, we define that for the availability property to be met, the distribution of $w$ across items should be statistically indistinguishable from that of $r$.

Note that some of these properties are \emph{inter-related}. For example, availability may be a necessary condition for comparability to be tested. Thus, it may be possible to design a fixed order in which these properties could be tested.

\section{Experimental Setup}
We describe the setup for testing if desiderata described in Section~\ref{sec:desiderata_list} are met in practice when relevance is inferred using searcher clicks.

\vspace*{-2mm}

\subsection{Data}
\label{sec:experiments_data}
Datasets in our experiments consist of a list of items which are to be ranked in response to a query. Each item is associated with some features and a relevance label. 
We assume that the ranking with highest-utility is one where items are arranged in decreasing order of these ground-truth relevance labels. Items also have labels denoting (protected) group membership.
We split each dataset in 70-10-20 proportions as train, validation, and test splits. The details of each dataset are below and summarized in Table~\ref{tab:ds_summary}. 
\looseness=-1

\begin{table}[tb!ht]
\centering
\small
 \begin{tabular}{cc|ccccc} 
 \toprule
 Dataset & $d$ & \multicolumn{5}{c}{Relevance Distribution, \% (4=highest)}\\
 & & 0 & 1 & 2 & 3 & 4 \\ 
 \midrule
\texttt{synth-normal} & 2 & 0.13 & 11.00 & 61.51 & 26.65 & 0.71\\
\texttt{synth-pareto} &  2 & 86.78 & 11.39 & 1.58 & 0.21 & 0.04\\
 \texttt{fairtrec} & 3 & 50.00 & 50.00 & - & - & -\\
 \bottomrule
 \end{tabular}
 \caption{Datasets statistics. $d$ denotes the number of features. Each column $i$ indicates the relative frequency of relevance grade $i$ in a given dataset (4: most relevant, 0: least relevant). Each dataset contains $50000$ samples}. 
 \label{tab:ds_summary}
 \vspace{-3mm}
\end{table}

\vspace*{-3mm}

\subsubsection{Synthetic Data} To experiment under controlled conditions, we synthesize two datasets from a graphical model using the synthetic data generation methodology proposed by ~\citet{yang2020causal}. In each case, the correlation Directed Acyclic Graph (DAG) consists of four attributes: continuous attributes for 2 features and a relevance label, and one binary group membership attribute. All three continuous attributes follow a Pareto or Normal distribution (see appendix for details).
After sampling, we discretize all relevance scores to 5 grades uniformly based on the value, similar to standard learning-to-rank setups.
We consider binary protected groups for all fairness analyses (e.g., men and women). 
We sample equal number of items from both groups.
We sample $N$=50,000 datapoints from both synthetic distributions.\looseness=-1
\looseness=-1

\textbf{Pareto distribution:} 
We sample relevance  from a Pareto distribution with $P(2.0, 1.0)$. This dataset is referred to as \texttt{synth-pareto}.\looseness=-1

\textbf{Normal distribution:} Relevance is sampled from a normal distribution with mean and standard deviation set to: $\mu=2, \sigma=1$ respectively.  This dataset is referred to as \texttt{synth-normal}.\looseness=-1

 \vspace*{-1mm}

\subsubsection{FairTREC 2021}
The FairTREC 2021~\cite{trec-fair-ranking-2021} dataset consists of Wikimedia articles ranked in response to text queries. In total, there are 50 train and 50 test queries. To align the setup of this dataset with our study, we choose a single query for analysis. We select a train-set query with atleast 25,000 number of relevance annotations available (train query ID 6). Additionally, only positive relevance annotations are available in the FairTREC dataset. We make the assumption that all the other Wikimedia articles in the corpus are not relevant to the query. Finally, we subsample the dataset to a similar size as the synthetic datasets to obtain a dataset with 50,000 items. Each item in this dataset thus consists of a query, article text, and a binary relevance judgment. To obtain a representation for ranking, we utilize a pre-trained checkpoint of a cross-encoder model trained on the \texttt{ms-marco}~\cite{craswell2021ms}\footnote{\url{https://huggingface.co/cross-encoder/ms-marco-TinyBERT-L-2-v2}} task. Finally, we embed the high-dimensional representations into three dimensions using Principal Component Analysis (PCA). This is the final ``item representation". Additionally, the geographical location metadata is utilized as the sensitive attribute for fair ranking. We binarize these categories into ``majority" location (Europe only) and all others (including no available metadata). This dataset is severely imbalanced in terms of group membership with a 90\% majority in group distribution.

\begin{figure*}[t!h]
    \begin{adjustbox}{width=6.5in}
    \centering
    \hfill
    \begin{subfigure}[b]{0.475\textwidth}   
        \centering 
        \includegraphics[width=.82\textwidth]{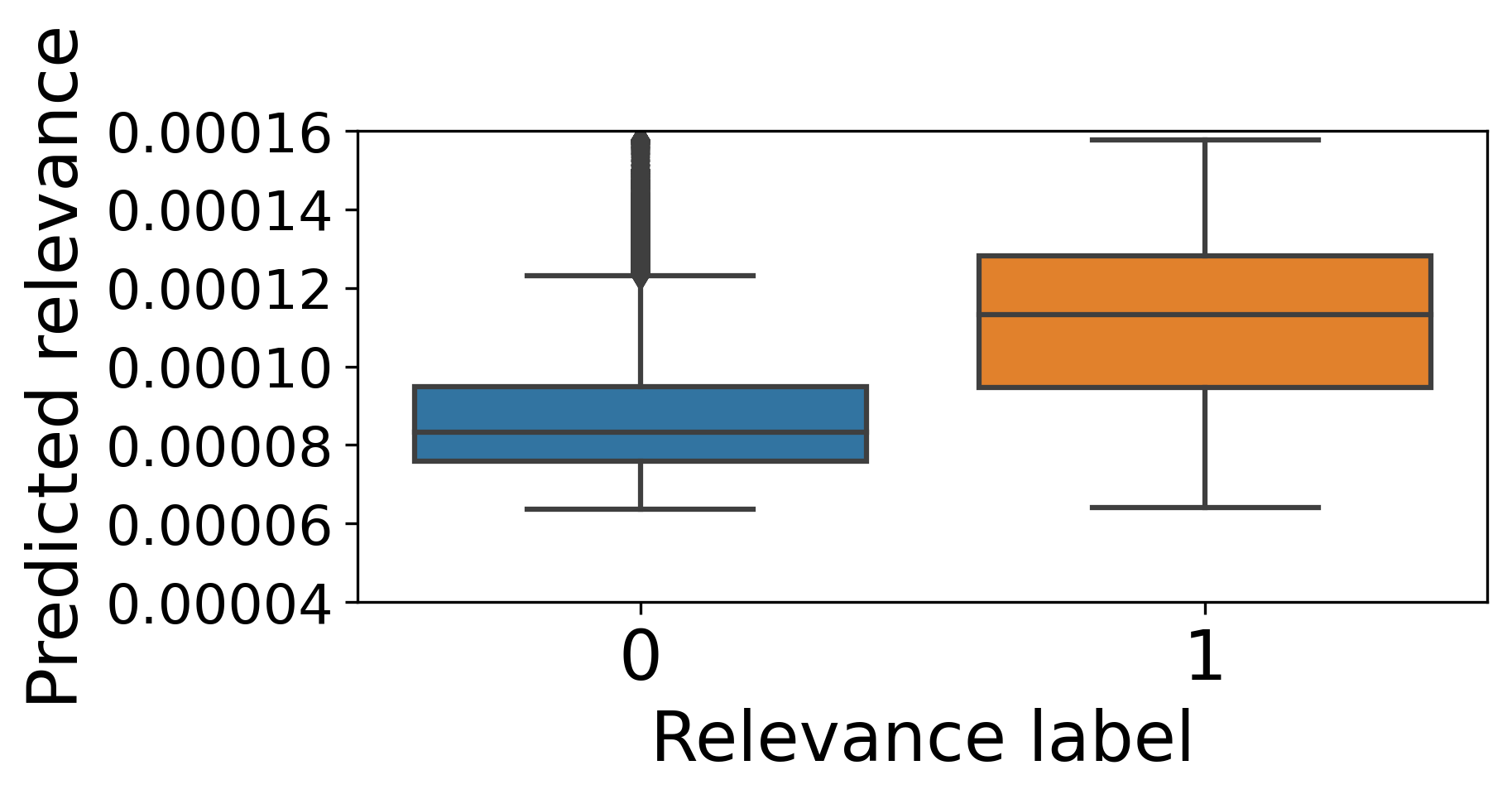}
        \caption{\texttt{fairtrec}\label{fig:fairtrec_corr}}%
    \end{subfigure}
    \begin{subfigure}[b]{0.475\textwidth}   
        \centering 
        \includegraphics[width=.82\textwidth]{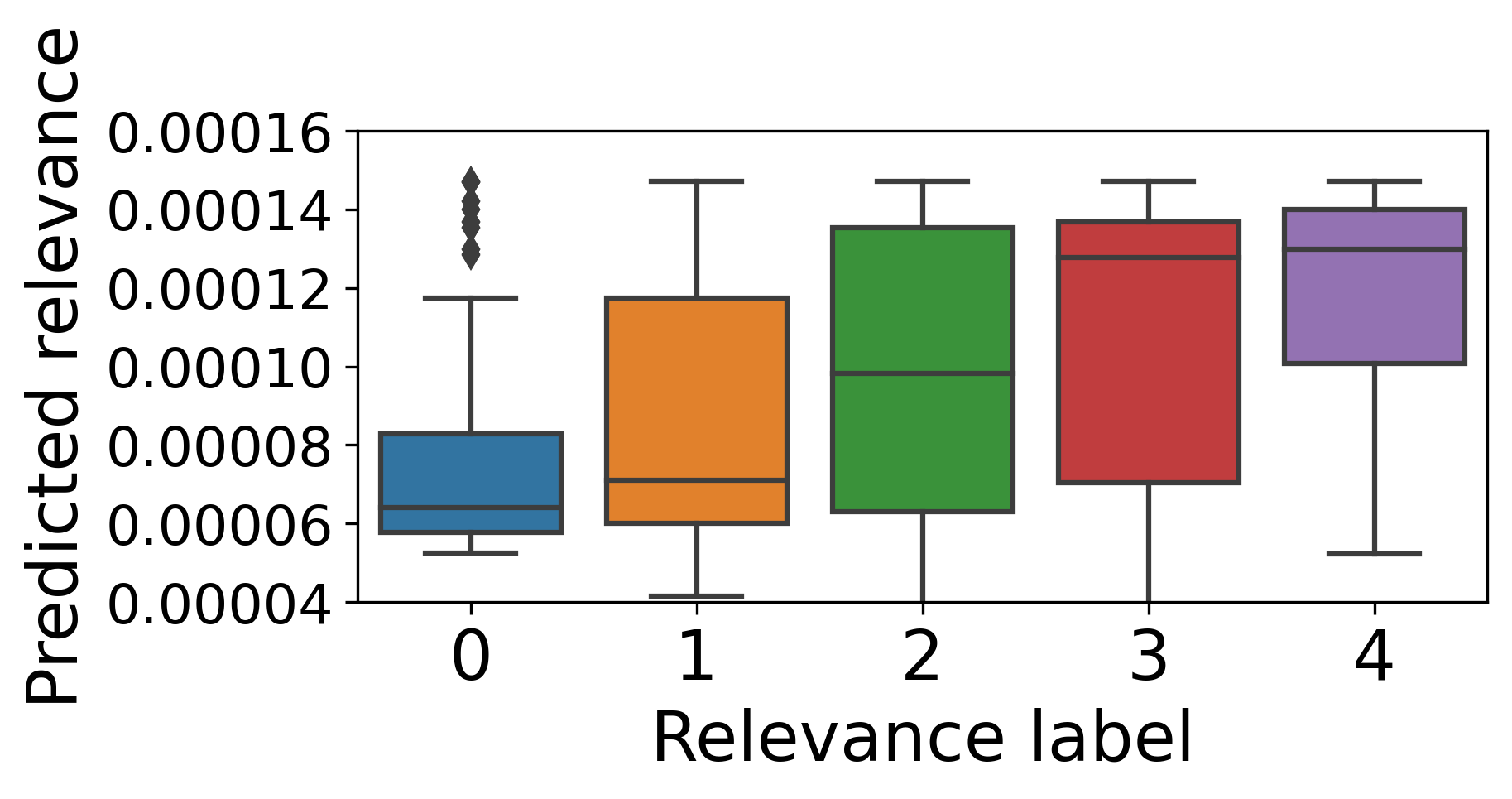}
        \caption{\texttt{synth-normal}\label{fig:normal_corr}}%
    \end{subfigure}
    \hfill
    \begin{subfigure}[b]{0.475\textwidth}  
        \centering 
        \includegraphics[width=.82\textwidth]{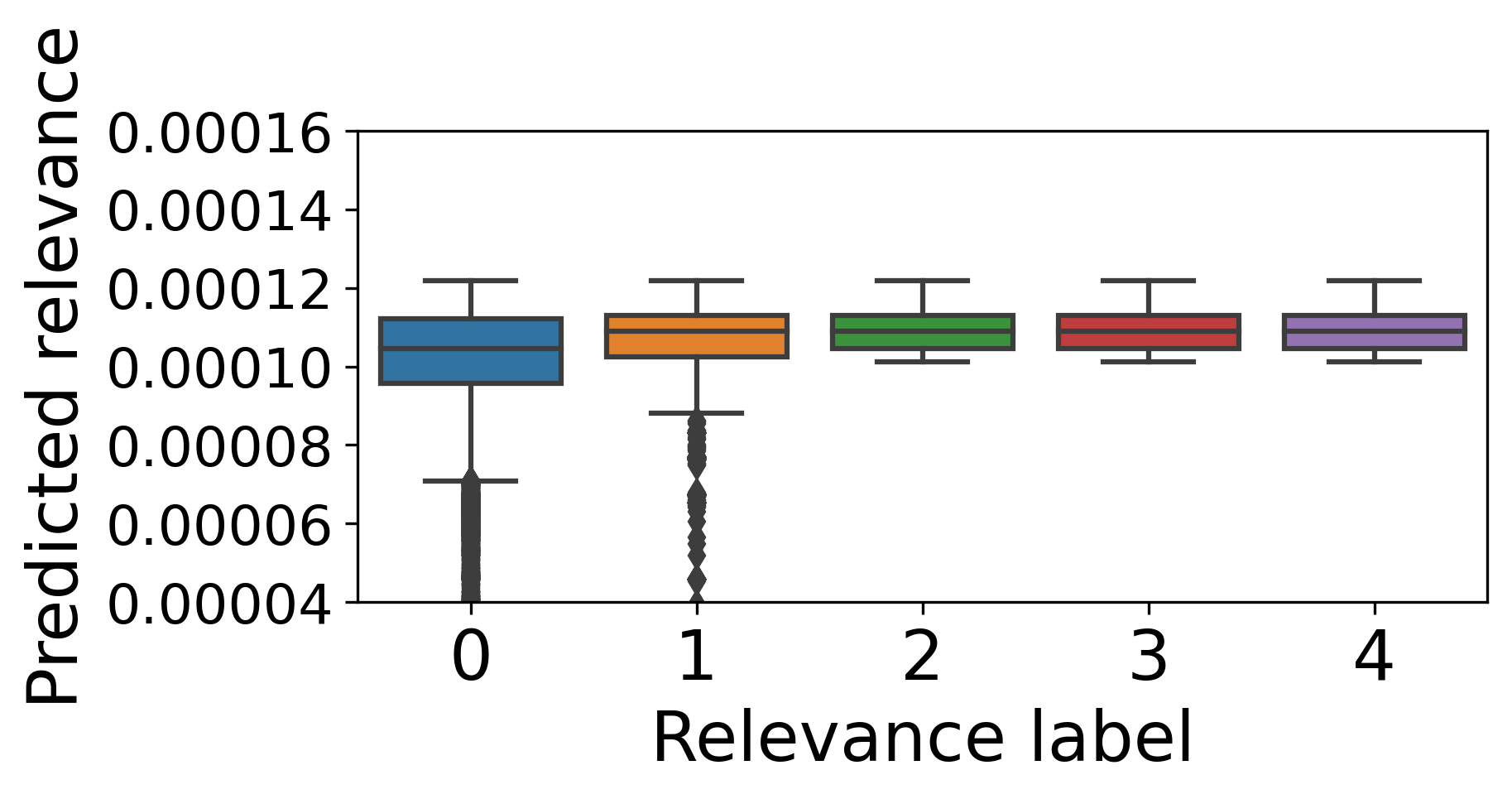}
        \caption{\texttt{synth-pareto}\label{fig:pareto_corr}}
    \end{subfigure}
    \end{adjustbox}
    \caption{Distributions of inferred vs true graded relevance scores. Items with high true relevance judgments must have high inferred relevance for the \emph{credibility} criterion to be met. We observe this to be true on average (i.e., ordering of medians is monotonic, though there is considerable overlap between inferred scores for different relevance grades). \looseness=-1}
    \label{fig:varying_corr}
\end{figure*}

\begin{figure*}[th!]
    \begin{adjustbox}{width=6.5in}
    \centering
    \hfill
    \begin{subfigure}[b]{0.31\textwidth}   
        \centering 
        \includegraphics[width=.85\textwidth]{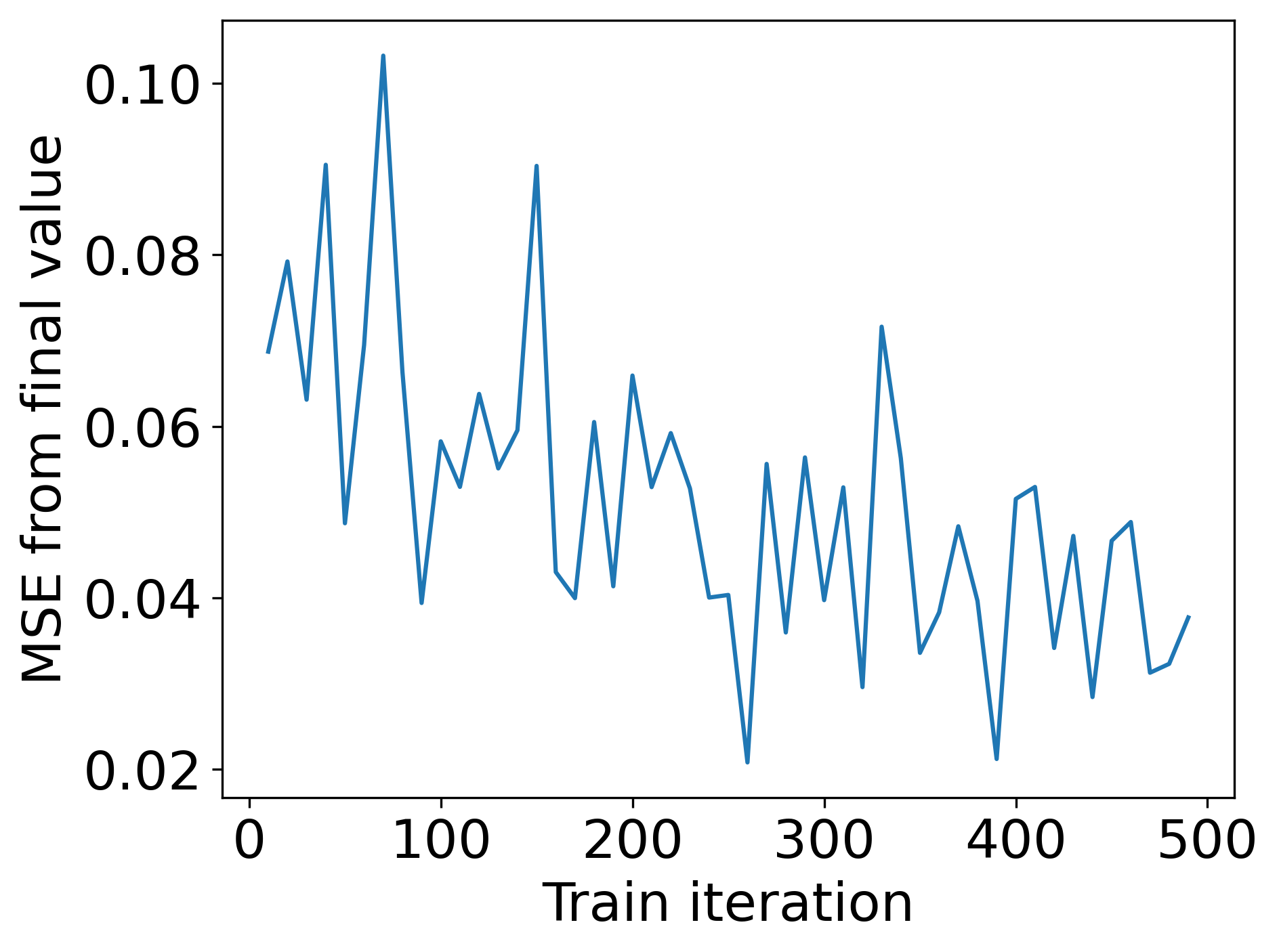}
        \caption{\texttt{fairtrec}\label{fig:fairtrec_consistency}}%
    \end{subfigure}
    \begin{subfigure}[b]{0.31\textwidth}   
        \centering 
        \includegraphics[width=.85\textwidth]{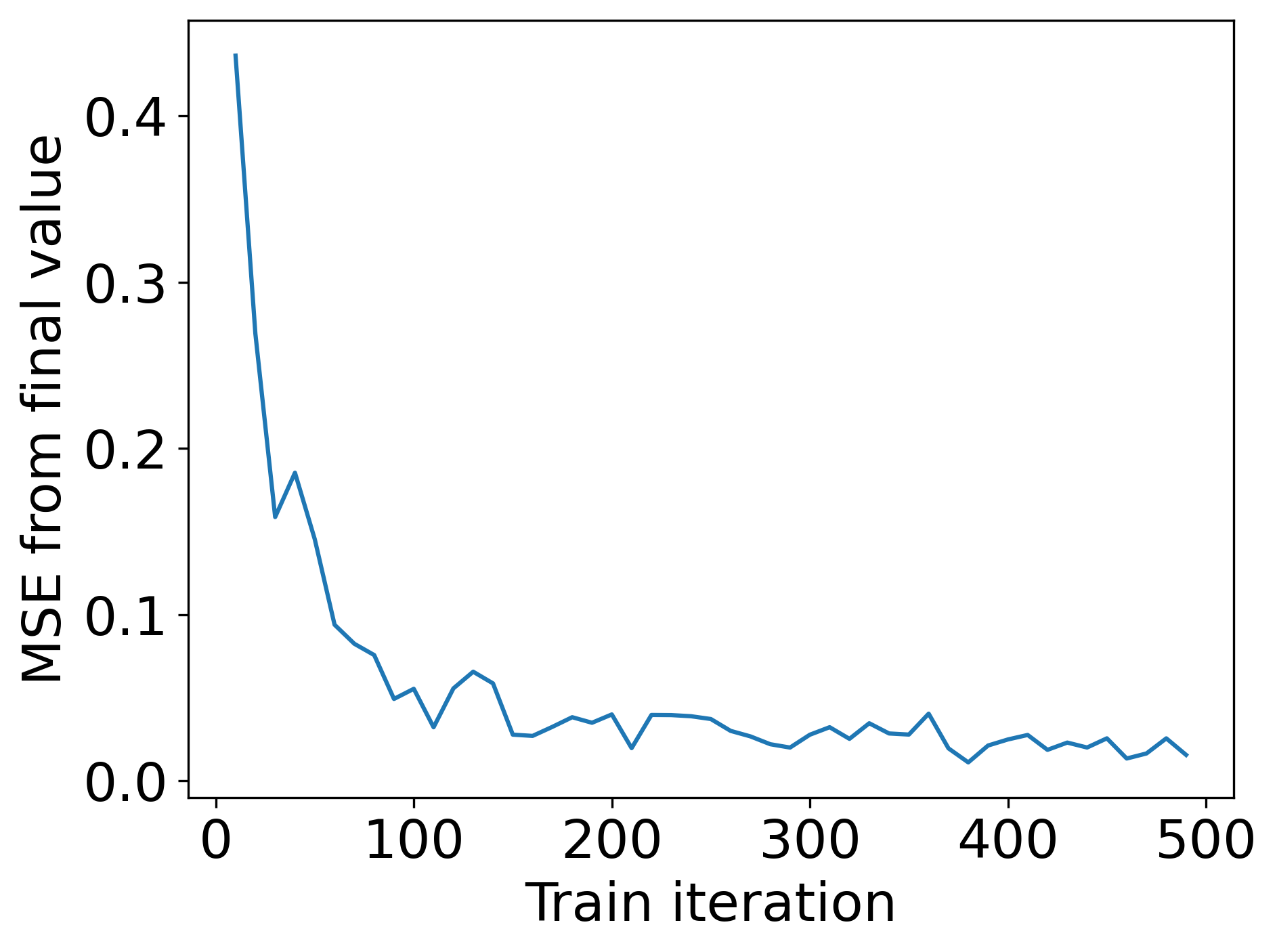}
        \caption{\texttt{synth-normal}\label{fig:normal_consistency}}%
    \end{subfigure}
    \hfill
    \begin{subfigure}[b]{0.31\textwidth}  
        \centering 
        \includegraphics[width=.85\textwidth]{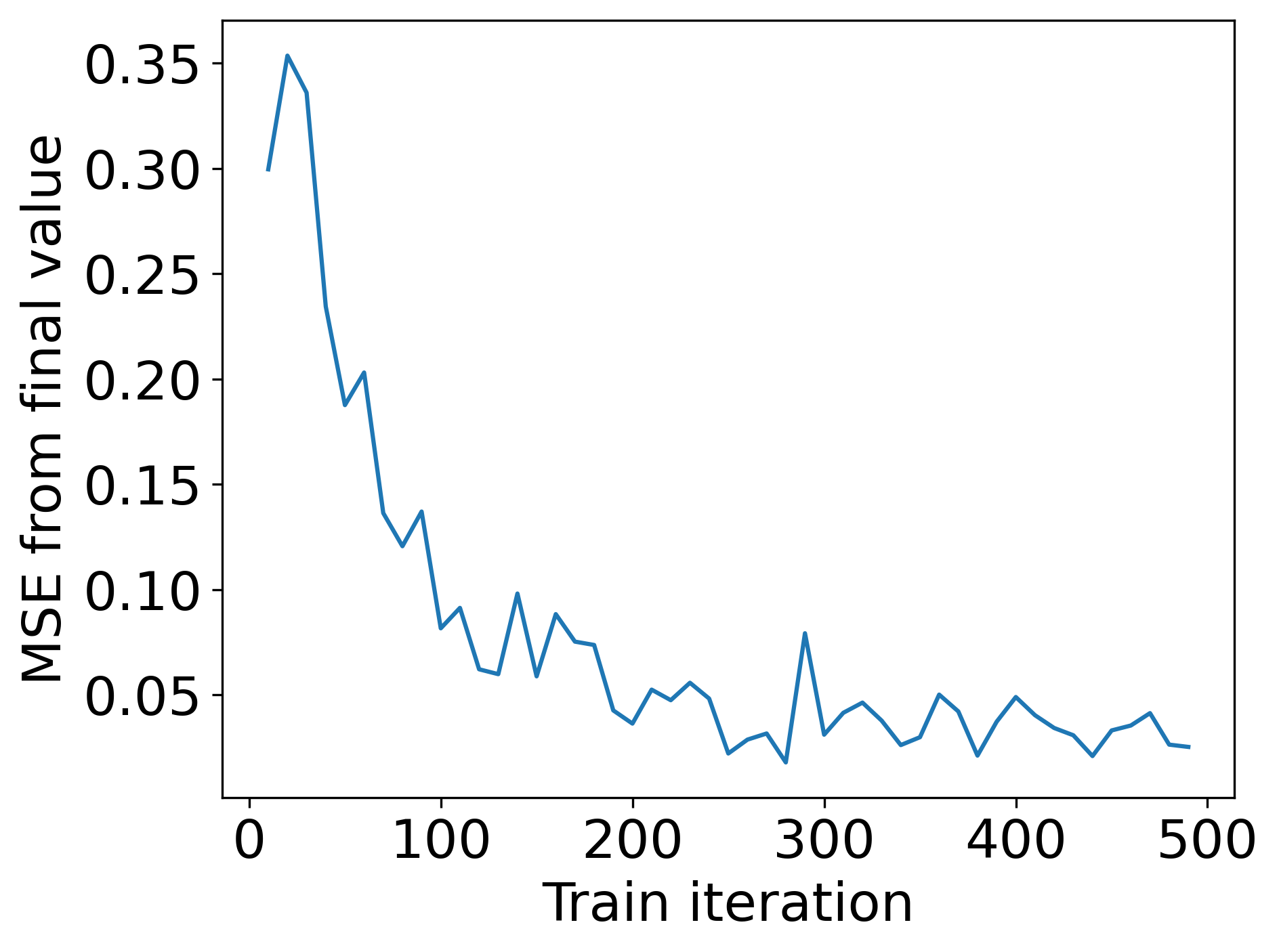}
        \caption{\texttt{synth-pareto}\label{fig:pareto_consistency}}
    \end{subfigure}
    \end{adjustbox}
    \caption{Consistency of predictions. Relevanceconsistently converges, as measured using  mean squared error (MSE). \looseness=-1}
    \label{fig:consistency}
\end{figure*}

\begin{figure*}[th!]
    \begin{adjustbox}{width=6.5in}
    \centering
    \hfill
    \begin{subfigure}[b]{0.31\textwidth}   
        \centering 
        \includegraphics[width=.81\textwidth]{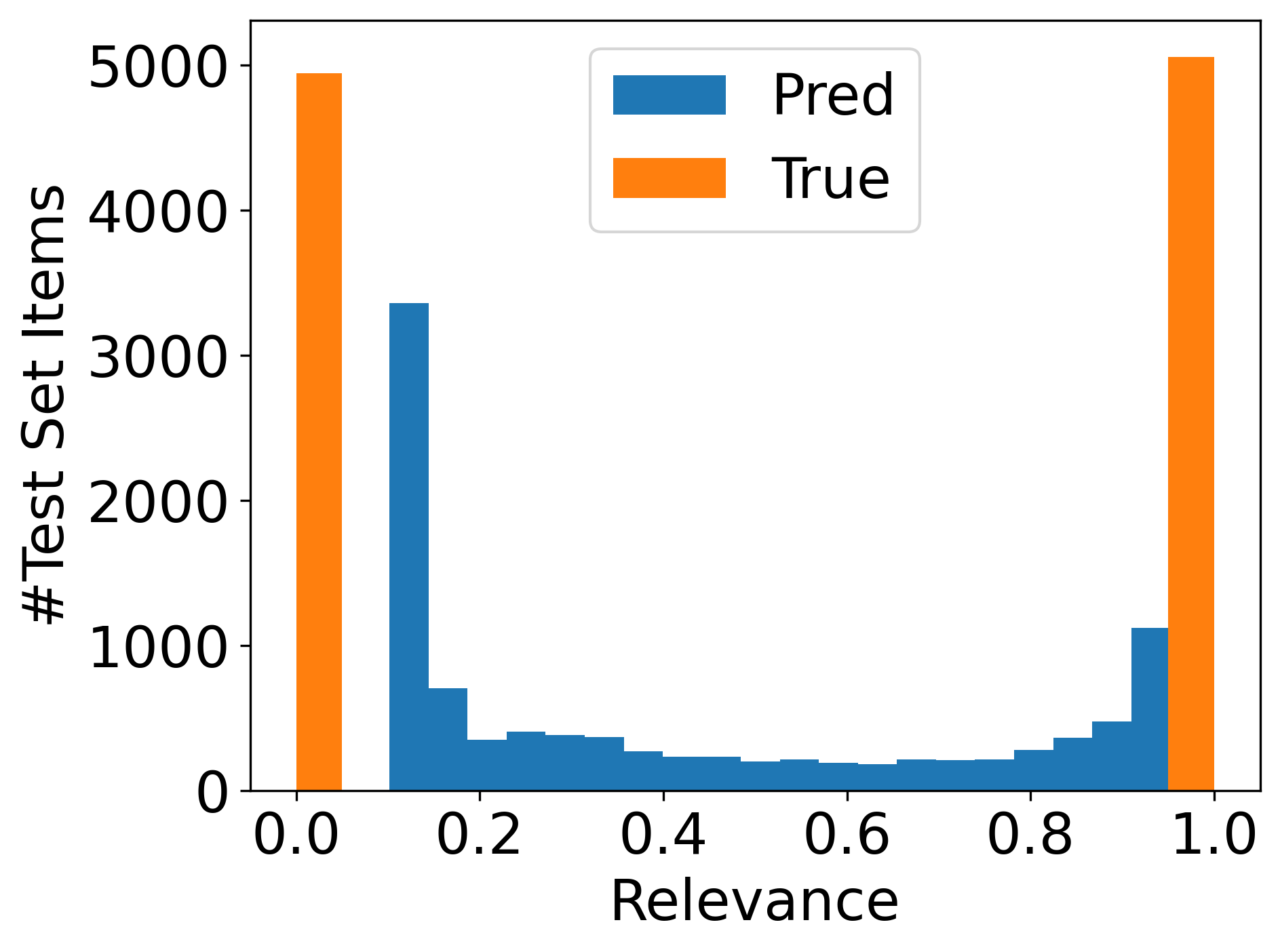}
        \caption{\texttt{fairtrec}\label{fig:fairtrec_hist}}%
    \end{subfigure}
    \begin{subfigure}[b]{0.31\textwidth}   
        \centering 
        \includegraphics[width=.81\textwidth]{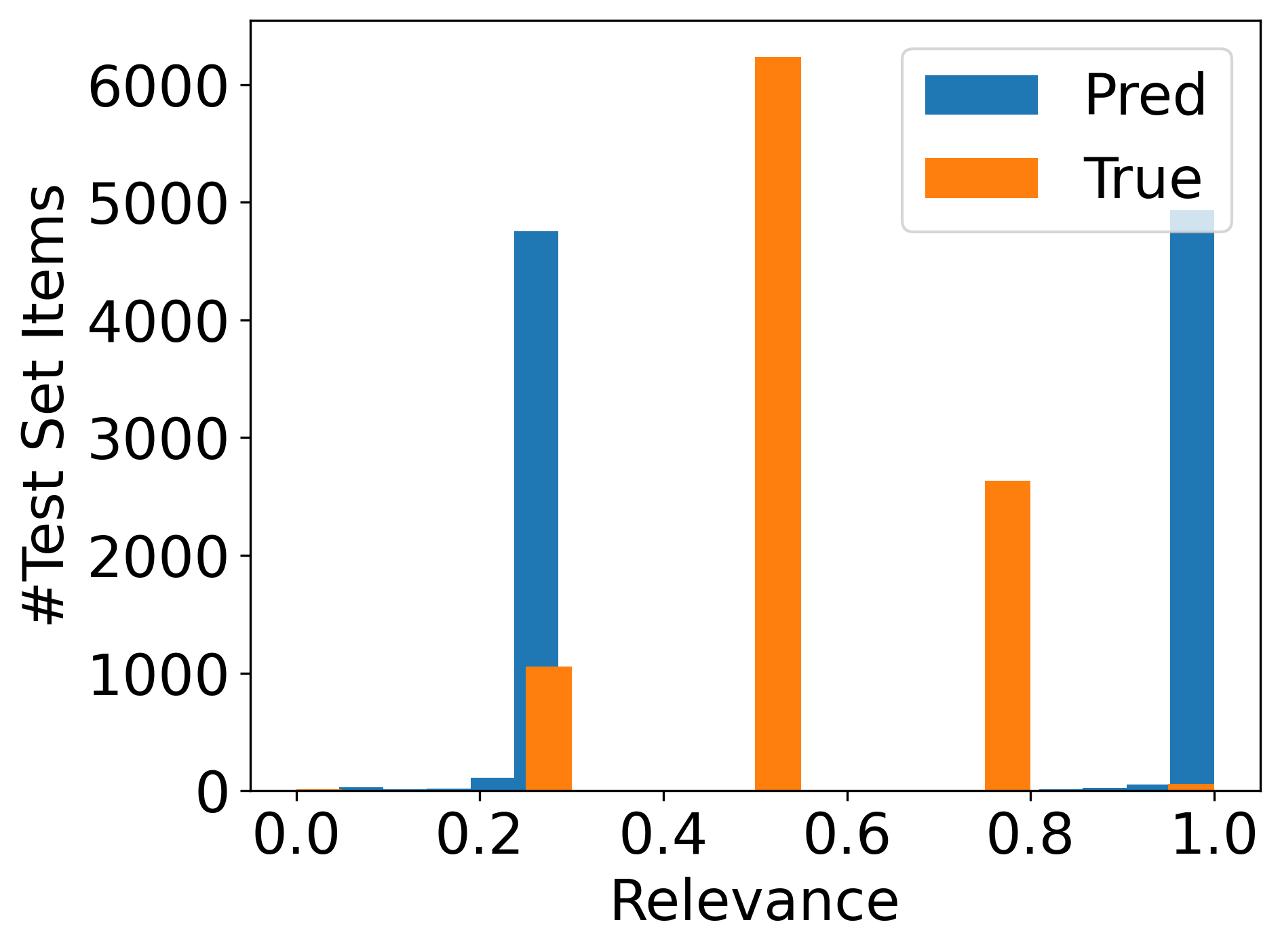}
        \caption{\texttt{synth-normal}\label{fig:normal_hist}}%
    \end{subfigure}
    \hfill
    \begin{subfigure}[b]{0.31\textwidth}  
        \centering 
        \includegraphics[width=.81\textwidth]{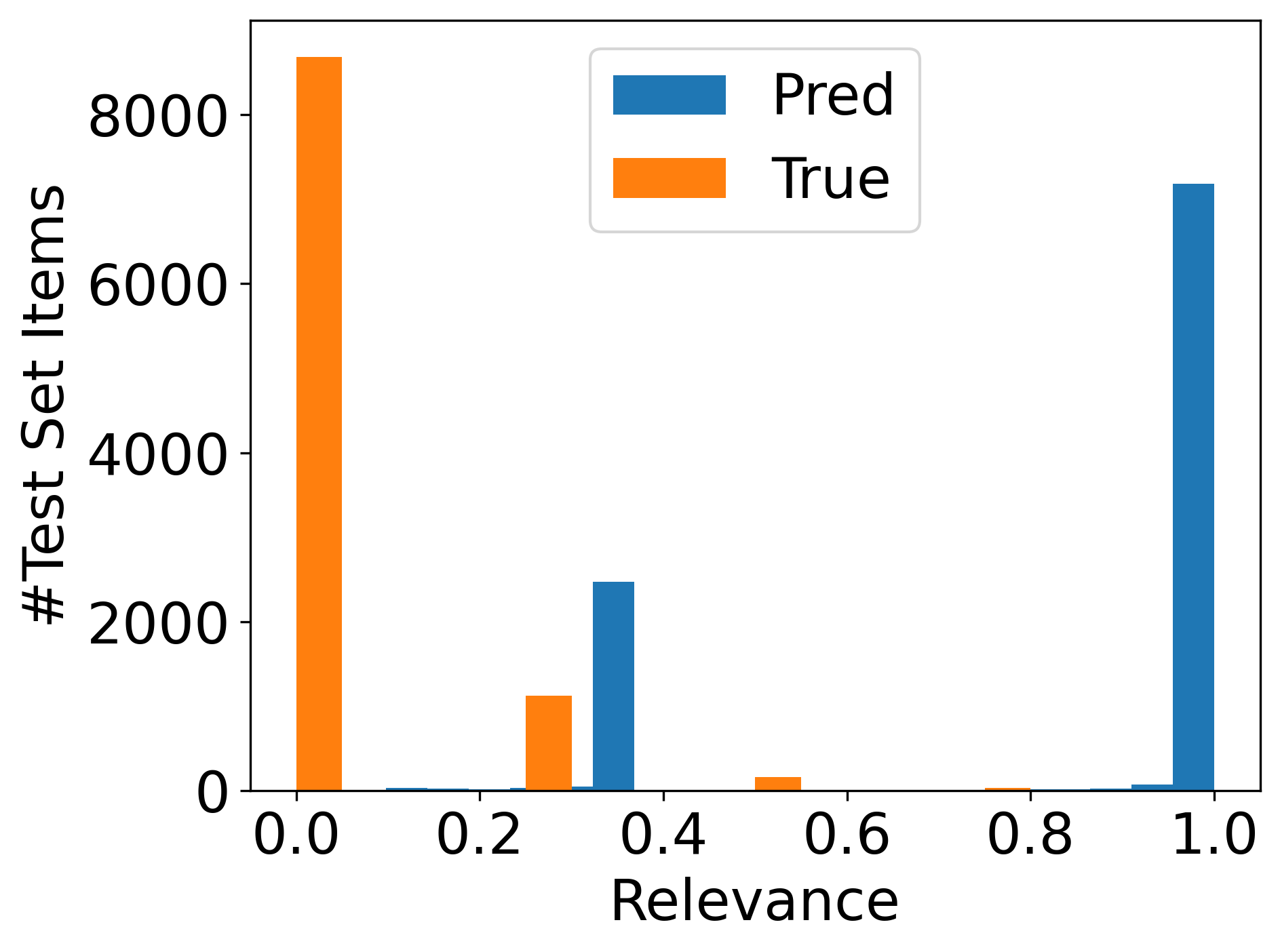}
        \caption{\texttt{synth-pareto}\label{fig:pareto_hist}}
    \end{subfigure}
    \end{adjustbox}
    \caption{Distributions of $[0,1]$-normalized inferred and true relevance scores: the inference process does not preserve distributions.
    }
    \label{fig:hist}
    \vspace{-3mm}
\end{figure*}

\vspace*{-2mm}

\subsection{Simulation Protocol}
We experiment in a {single-query} setup without personalization. In each round, we generate synthetic clicks to simulate a set of users clicking on items based on the position-biased examination probability and relevance of each item to be ranked, using a position-biased click model (as specified in Sec.~\ref{sec:rel_from_clicks}; with a positive click probability of $1$, negative click probability of $0.1$, and maximum relevance grade of $4$). Then, the average click through-rate of each item is accumulated to compute its propensity and the ranking model is optimized with inverse-propensity weighted optimization (Sec.~\ref{sec:unbiased_ips}). In our experiments, we use pre-computed propensity weights using result randomization (i.e., a pre-trained propensity model) for the position-biased click model\footnote{From \href{https://github.com/ULTR-Community/ULTRA_pytorch/blob/main/example/PropensityEstimator/randomized_pbm_0.1_1.0_4_1.0.json}{the ULTRA-pytorch simulation repository}.} while training the ranking model. The metric optimized in each step is the list-wise ranking loss with a selection-bias cutoff (set to 10). The final ranking system output is a list of items, ranked in decreasing order of expected utility to users. This is repeated for 500 iterations in total during training, and performance is evaluated, with models checkpointed every 50 iterations. Model loss is optimized with stochastic gradient descent with a learning rate of $0.01$.\footnote{We validated that an Adagrad optimizer produced similar train loss.} Performance on the test set is evaluated with the best model checkpoint (based on validation performance). This process is repeated with 10 random seeds, and averaged for all metrics. Note that there are several sources of randomness in this setup, including the items that are clicked, the initialization of parameters in the ranking model, and sampling of data batches during training. 
\looseness=-1

\looseness=-1

\vspace*{-1mm}

\section{Experiments and Results}
\label{sec:exps_results}
In this section, we seek to answer the question: To what extent can click-based relevance be used to guide fair exposure allocation? 
\looseness=-1

\vspace*{-2mm}

\subsection{Studying Properties of Relevance}
\label{sec:sec_studying_properties}
We empirically test if the desired criteria for relevance are met in practice on three datasets. We assume that the true relevance labels are proportional to true worthiness, and we refer to ground truth relevance judgment labels as true relevance, and inferred relevance labels as predicted relevance. \looseness=-1
We note that both the tests for operationalizing the formal criteria (see Section~\ref{sec:desiderata_list}) and the thresholds for meeting these criteria are themselves choices. For example, credibility can be measured using the mean deviation between the inferred relevance scores and true relevance scores for point-estimates, or by measuring the overall Spearman correlation between them (which may be closer to the rank ordering definition). Further, to say that a criterion is met, one may set acceptance thresholds, e.g., that a given criterion might be met only if a specific correlation is greater than $0.5$. Thus we emphasize that meeting these criteria may be a continuum, not a binary outcome. Summary of findings is in Table~\ref{tab:rel_properties_summary}.
\looseness=-1

\vspace*{-2mm}

\subsubsection{Credibility}
\label{sec:relevance_reliability}
We study if relevance inferred from clicks under this prototypical setup is credible. Specifically, we test if the top-items across multiple runs of the learning-to-rank system (i.e., different seeds) correspond to top items (i.e., the ordering is retained). We construct boxplots showing the distribution of relevance scores (with full list size being the size of the test set; all softmax scores within a given ranking for ease of interpretation) for each relevance judgment label grade. Ideally, we expect that items with higher relevance grades would occupy lower ranks by having high inferred relevance scores. We assert that the relevance scores are not credible if the population medians corresponding to different relevance grades are not significantly different at a level of $0.05$.
\looseness=-1

\textbf{Results.}  From Figure~\ref{fig:varying_corr}, we observe that for increasing values of true relevance judgments, the medians of the inferred relevance scores for items are monotonically increasing for all the datasets. With a Kruskal Wallis-H test, we observe that these medians are significantly different ($p<0.001$). This lets us conclude that the relevance scores are indeed reliable across runs for all three datasets. However, post-hoc tests may be required to test if medians of specific groups (e.g, those with relevance grades of $2$ vs $3$) are significantly different.

\vspace*{-2.2mm}

\subsubsection{Consistency}
\label{sec:relevance_consistency}
We measure the mean squared error between the predicted relevance scores at iteration $i$ and the final iteration (i.e., $i=500$) for $i=10,20,30 \ldots 490$. Note that we use the logit-score predictions, before the \emph{softmax} operation. These scores are computed on the validation rather than test set (since test set is unseen throughout training). Then, we assess if this mean squared error converges. Based on the definition for consistency (see Section~\ref{sec:desiderata_list}), we need to choose a $\epsilon$ value to assess consistency. Here, we set $\epsilon=0.1$ \looseness=-1

\textbf{Results.} We observe that the mean squared error converges to the final value for all datasets, though with some noise in \texttt{fairtrec}. As a result, the consistency criterion is satisfied all datasets on the validation set.
(Consistency also met for datasets on the train set.)
\looseness=-1

\vspace*{-2.2mm}
\subsubsection{Stability}
\label{sec:relevance_stability}

We measure the deviation in inferred relevance values across multiple runs. We compute the average standard deviation in relevance measurement of a specific item across the ten random seeds, and average this value across all items. To ensure that all values can be compared across datasets, we scale predicted scores for each item in given list by subtracting the mean and dividing by standard deviation, all derived from the predictions for all items in that list.\looseness=-1

\textbf{Results.}  Among the synthetic datasets, we find that the mean variation is is between $0.17 - 0.18$. Since all relevance scores have been normalized with unit-variance, this corresponds to less than one standard deviation of the full relevance distribution. The results for the \texttt{fairtrec} dataset show similar trends (variances scores $\sim$0.41). As a result, we conclude that the relevance scores are reasonably stable for all datasets.
\looseness=-1

\vspace*{-2.2mm}

\subsubsection{Comparability}
\label{sec:relevance_comparability}
We compute two metrics to assess the comparability criterion: (1) the Spearman correlation between predicted and true relevance scores for all items, (2) the ratio between average relevance scores for groups for true and predicted relevance scores. We assess that the criterion is met if the correlation is $\geq$ 0.3 from (1) and the difference in ratios from (2) is not more than $0.05$. Note that (1) is  closely tied to individual fairness since it considers prediction for all items. On the other hand, (2) is directly consequential to group-level exposure fairness: specifically, exposure is allocated in proportion to expected group relevance (either predicted or true) as per several fair ranking formulations~\cite{biega2018equity,joachims2007evaluating,raj2022measuring}.

\textbf{Results.} Spearman correlation between true and predicted relevance scores is 0.09 (0.09 for subgroups) for \texttt{synth-normal}. For \texttt{synth-pareto}, the Spearman correlation between true and predicted relevance scores is 0.23 (0.20, 0.26 for female and male subgroup respectively). This indicates that the comparability criterion of relevance scores is better for one subgroup here. In contrast, the trends observed on the \texttt{fairtrec} dataset are different:  the correlation is 0.27 for all items (subgroup correlations of 0.27 and 0.29). \looseness=-1  

The ratio between normalized average relevance for subgroups using true relevance scores are close to 1 for both synthetic datasets (by construction). Ratios between average inferred group-relevance scores are also close to 1.00 for \texttt{synth-normal} and \texttt{synth-pareto} respectively. The difference for \texttt{fairtrec} is similar, with the two ratios being 1.04 and 1.02 with true and inferred relevance respectively. Thus, the target for exposure ratios may change slightly on using true vs predicted relevance scores. \looseness=-1 
In this case, as per the criteria set, the comparability criterion is not satisfied for all datasets. The comparability criterion related to group-fairness are met on all datasets. \looseness=-1  

\vspace*{-2mm}

\subsubsection{Availability}
\label{sec:relevance_availability}
We analyze the distribution of inferred relevance scores and compare them to the distribution of true relevance labels. 
\textbf{Results.}  By design, since a machine learning model can make predictions for all items, the continuous relevance score predictions are available for all items. However, when we visualize the histogram of predicted and true relevance (plots in Figure ~\ref{fig:hist}), we observe that the nature of distributions is not retained. We computed the two-sample Kolmogorov-Smirnov test~\cite{massey1951kolmogorov} for goodness of fit with null hypothesis being that the two samples -- true and inferred relevance scores -- arise from the same distribution. We find that the null hypothesis can be rejected in all cases with $p\ll 0.001$. As a result, we conclude that the availability criterion is not satisfied for at least subset of ranked items. This makes as sense as available ``worthiness" judgements here are graded, while predictions are continuous.\looseness=-1

\begin{table}[tb!h]
\begin{adjustbox}{width=\linewidth}

\begin{tabular}{lccccc}
\toprule
Dataset &   Credibility  & Consistency & Stability & Comparability & Availability  \\
\midrule
\texttt{fairtrec}  &  \cmark & \cmark   & \cmark & \xmark & \xmark \\
\texttt{synth-normal} &  \cmark & \cmark & \cmark & \xmark & \xmark  \\
\texttt{synth-pareto} &  \cmark & \cmark  & \cmark & \xmark & \xmark  \\
\bottomrule
\end{tabular}
\end{adjustbox}
    \caption{Empirical tests of relevance score desiderata on three datasets. In this table, we summarize whether the five desired criteria are empirically met in the datasets. \looseness=-1 \label{tab:rel_properties_summary} %
    \vspace{-5.5mm}
    }
\end{table}

\vspace*{-2mm}
\subsection{Impact of Violating Criteria in Practice}
\begin{table*}[ht!]
    \begin{subtable}{.37\linewidth}
      \centering
            \begin{tabular}{l|cc}
            \toprule
            & \multicolumn{2}{c}{Individual Fairness}\\
            \cline{2-3}\\
            Dataset &   True  &  Predicted    \\
            \midrule
\texttt{fairtrec} & 19.703 $\pm$ 0.01  & 19.776 $\pm$ 0.01$^{*}$  \\
\texttt{synth-normal} & 19.971 $\pm$ 0.03  &  19.966 $\pm$ 0.03$^{*}$ \\
\texttt{synth-pareto} & 19.523 $\pm$ 0.03 &  19.949 $\pm$ 0.03$^{*}$\\
\bottomrule
            \end{tabular}
            \caption{\label{tab:ind_fair}}
    \end{subtable}%
    \begin{subtable}{.23\linewidth}
      \centering
         \begin{tabular}{cc}
            \toprule
            \multicolumn{2}{c}{Exposure Fairness}\\
            \cline{1-2}\\
            True  &  Predicted    \\

            \midrule
  $0.942 \pm 1.03$ & $0.953 \pm 1.03^{*}$\\
  $3.420 \pm 0.31$ & $3.407 \pm 0.31^{*}$\\
  $2.180 \pm 1.01$ & $2.210 \pm 1.01^{*}$ \\
\bottomrule
            \end{tabular}
             \caption{\label{tab:exp_fair}}

    \end{subtable} 
    \begin{subtable}{.19\linewidth}
      \centering
         \begin{tabular}{cc}
            \toprule
            \multicolumn{2}{c}{Exposure fairness + intervention}\\
            \cline{1-2}\\
            DetConstSort  &  DetCons    \\

            \midrule
$1.023 \pm 1.11^{*}$ & $1.020 \pm 0.93$\\
$3.131 \pm 0.34^{*}$ & $1.420 \pm 0.18^{*}$\\ 
$2.154 \pm 0.91^{ }$ & $1.901 \pm 0.69$ \\

\bottomrule
            \end{tabular}
             \caption{\label{tab:fairness_intervention}}

    \end{subtable} 
    \caption{(a) Individual fairness, (b) Group Exposure fairness measured on the test set, averaged across 10 runs, using true vs predicted relevance scores for the same set of rankings, (c) Impact of group-level fairness interventions. We observe that when inferred relevance scores are used, fairness metric values are significantly different (paired Wilcoxon test; $p<0.05$; not significant with unpaired tests), though the size of the difference is often small. The deviation reported for individual fairness measurements corresponds to variance across items.
    (c) displays fairness with predicted relevance after unfairness mitigation algorithms intervene.\looseness=-1}
    \vspace{-5.3mm}
\end{table*}

While relevance is the construct measured in ranking systems, the user-facing output is a ranked list of items (which is a function of the inferred relevance). Hence, system evaluations form an important part of optimizing ranking systems. We evaluate ranking systems along two axes: (1) quality or utility of ranking, (2) fairness of ranking. We measure the quality of rankings using NDCG@10 score for all datasets, and obtain scores of $0.90$ (near perfect score)\footnote{We use pre-trained ranking models for embeddings; modifying this led to similar results.}, $0.29$, and $0.26$ for \texttt{fairtrec}, \texttt{synth-pareto}, and \texttt{synth-normal} respectively.\looseness=-1 

To test if violating the criteria matters in practice, we measure exposure fairness using the metrics defined in Sec.~\ref{sec:related_work_fairness}, all computed either using true, \emph{graded} or inferred \emph{continuous} relevance scores on the test set. From Table~\ref{tab:exp_fair} and Table~\ref{tab:ind_fair}, we observe that fairness assessments may vary depending on whether true or inferred relevance scores are used, and the size of difference is often dataset-dependent. In particular, the difference does not seem particularly stark for the synthetic datasets. However, on computing Wilcoxon signed-rank tests paired by the random seed (which influences the stochastic initialization of the prediction model), the exposure fairness metrics are significantly different for all datasets. Similar trends are observed for individual fairness: the difference is significantly different, where pairing is performed at item-level (i.e., 10000 items). We observe that these results are sensitive to normalization of relevance scores.\looseness=-1 

Thus, fairness metrics vary depending on whether true or predicted relevance scores are used. Notably, the comparability criterion is also violated (especially for individual fairness) on all three datasets  (Section~\ref{sec:sec_studying_properties}). Analyzing the degree to which a criterion is violated may yield interesting insights. For example, the difference in ratios of group-level relevance assessed during comparability evaluation is generally small for the synthetic datasets (Sec.~\ref{sec:relevance_comparability}), as is the difference in group-level exposure fairness (Table~\ref{tab:exp_fair}). Thus, the difference between fairness metrics using true and predicted relevance may be closely tied to the comparability criterion of predicted scores. \looseness=-1

\vspace*{-1.8mm}

\subsection{%
Mitigating Measurement Issues}

In  this section, we highlight examples of conditions and interventions that might mitigate relevance measurement issues. 
\looseness=-1

\subsubsection{Mechanisms: Fairness Interventions}
We study a post-hoc fairness processing setup with two algorithms proposed by \citet{geyik2019fairness}. Both fair re-ranking algorithms  (DetCons, DetConstSort) rearrange the top-k items in a ranked list such that the distribution of groups matches a user-specified distribution in a greedy manner (see ~\citet{geyik2019fairness} for details). With a group-fairness lens, we set the desired proportion of a group among the top-k items to be the average relevance of items in that group divided by the sum of average relevance of items for all groups.\looseness=-1

We study the impact of these interventions on estimated group-level exposure fairness.
Do the interventions \emph{mitigate} the measurement issues in fairness? From Table~\ref{tab:fairness_intervention}, the fairness interventions are successful in reducing system unfairness (ratios are closer to $1$ than pre-intervention scores in Table~\ref{tab:exp_fair}). In some of the six cases (two algorithms, three datasets), the re-ranking algorithm modifies the system fairness to be indistinguishable from the pre-intervention value with true relevance scores (e.g., the \texttt{fairtrec}+DetCons experiment, and the \texttt{synth-pareto}+DetConstSort experiment). We observe that results are sensitive to the top ``k" chosen for re-ranking.
Thus, fairness interventions may be a mechanism to reduce the error in fairness assessment, but impact may vary based on the algorithm.\looseness=-1

\vspace*{-1.5mm}

\subsubsection{Data: Imbalanced Groups}
\label{sec:imbalance}
During synthetic data generation (Sec.~\ref{sec:experiments_data}), we ensured equal group sample sizes.
While class imbalance is not explicitly a component of the measurement model of relevance, it is often an artefact of ranking system design, and therefore could affect system fairness evaluation. Here, we consider various levels of data imbalance in the datasets, and study the resultant effects on exposure fairness---i.e., does imbalanced training data mitigate or hide the effects of relevance desiderata violations? We simulate imbalance ranging from 50-50\% (i.e., no imbalance) to 90-10\% (i.e., high imblance) by subsampling the synthetic datasets to 25,000 points. \looseness=-1

For both datasets, we observe that the degree of imbalance impacts the difference between exposure fairness metrics as computed with true and predicted relevance: the difference ranges from about $-0.10$ to $0.12$, and about $-0.03$ to $0.00$ for \texttt{synth-pareto} and \texttt{synth-normal} respectively. A negative value here implies that the exposure fairness assessed using the predicted relevance score has a lower value than that using true relevance. When fairness measurements here are greater than $1$ (true for most of imbalance ratios here), this implies that the system appears more fair when predicted relevance is used.  Thus, at specific imbalance ratios, the system may appear \emph{more fair} than it is.  In summary, group sample size imbalance may impact the inferred relevance scores and thus system fairness. \looseness=-1

\vspace*{-3.9mm}
\section{Future Perspectives for IR}
 A turn towards examining fairness \cite{castillo2019fairness} and equity \cite{biega2018equity} in ranking has inspired new research agendas in IR. Meanwhile attending to the two-sided nature of ranking platforms has  broadened the desiderata of fair ranking systems \cite{patro2022fair,ekstrand2022fairness,phillips2023algorithmic}. 
Past works propose interventions to fairly allocate opportunities and avoid exacerbating social biases based on some latent, complex notion of worthiness~\cite{shakespeare2020exploring,wachs2017men,mehrotra2018towards,sapiezynski2019quantifying,ferraro2021fair,biega2018equity}. Ultimately these interventions rely on \emph{relevance} as a convenience proxy to allocate exposure. 
Our work points to a pressing need for information retrieval to reengage with research on relevance in the context of fair ranking. 
\looseness=-1

\textbf{Critically assessing limitations of relevance as a proxy for worthiness.}
Relevance can be justified as a proxy for worthiness; however, this justification must be done by meaningfully establishing the validity and reliability of relevance in a given setting. For the goal of fair ranking, this requires showing that the properties of the relevance scores and resulting rankings are aligned with the intended fairness goals. In our paper, we demonstrate an example of validating several such properties. Concrete use cases (e.g., hiring, marketplaces) and intended normative goals should guide the acceptable limitations of relevance. \looseness=-1

The properties proposed in this paper were conceptualized based on extensive discussions between the authors of this paper using interdisciplinary lenses of machine learning, sociotechnical systems, measurement theory, and information retrieval. Based on the application at hand, some of these properties may be more important than others. We also emphasize that additional properties may still need to be identified and tested for in various use-cases. Still, we believe that the proposed framework of elucidation followed by examination of properties offers a step towards assessing if a specific proxy (such as relevance) can be reliably used as a target for fair ranking. In concurrent work with an alternate view, ~\citet{schumacher2022properties} derive desired properties for group fairness metrics and highlight similar nuances of metrics that rely on relevance, further validating our findings. In contrast, we derive desiderata for \emph{relevance} to guide fair exposure allocation. \looseness=-1

\textbf{Defining `worthiness' for different ranking application domains.}
Different theoretical conceptualizations of worthiness will correspond to different normative goals and theories of justice. For instance, one could distinguish between the fairness of process (where the worthiness score might encode the merit of the ranked subjects) and the fairness of consequence (where the worthiness score might correspond to the utility to a selected stakeholder of a subject being top-ranked). The design goals of a system will also reflect their application domains: a marketplace of online sellers might think differently about worthiness than a hiring platform. Precisely connecting different conceptualizations of worthiness to how they are operationalized in fair ranking systems will aid researchers, developers, and auditors to enhance system equity. \looseness=-1

\textbf{New methods for obtaining worthiness scores.}
Beyond the need for new definitions, this paper points to the importance of considering how worthiness is operationalized in the context of fair ranking as well as revealing the need for new operationalizations. 
Potential approaches might vary from finding better proxies, developing new methods for direct worthiness crowdsourcing, finding new ways of accounting for annotator biases, calibrating predictions of relevance from browsing models, or proposing new approaches that go beyond existing IR relevance methodologies.

\section{Conclusions}
Ranking systems mediate opportunity in a variety of high impact settings~\cite{joachims2017accurately}. In this paper, we leveraged an interdisciplinary perspective to show how fair ranking systems may fail to engage with the intended goals of fair ranking, relying instead on a tenuous assumption of relevance as worthiness.
Drawing on domain knowledge about relevance-based fair ranking, we derived a set of criteria that relevance must satisfy in order to be a \emph{valid} and \emph{reliable} target to guide fair exposure allocation: credibility, consistency, stability, comparability, and availability.  Using click-based relevance as a case study, we tested if these criteria can be empirically met on three datasets. We observe that a subset of these criteria may be unfulfilled in practice. 
\looseness=-1
Beyond click-based relevance, similar criteria will have to be tested for other forms of relevance estimation (e.g., crowdsourced relevance judgments). The contributions of our work are in establishing a measurement theory-based framework for thinking about the role of relevance in fair exposure allocation settings. More broadly, our work highlights the need for novel approaches to generate and collect relevance scores in a valid, reliable manner in fair ranking settings. Together, this reveals  several open problems in relevance measurement at scale for fair exposure allocation. \looseness=-1

\begin{acks}
This paper is part of the FINDHR (Fairness and Intersectional Non-Discrimination in Human Recommendation) project that received funding from the European Union's Horizon Europe research and innovation program under grant agreement No 101070212.
\end{acks}

\clearpage
\bibliographystyle{ACM-Reference-Format}
\bibliography{sample-base}

\clearpage
\appendix

\section{Synthetic Data Generation Graph}
\begin{figure}[h]
\centering
\begin{subfigure}{\linewidth}
  \centering
  \includegraphics[width=0.6\linewidth]{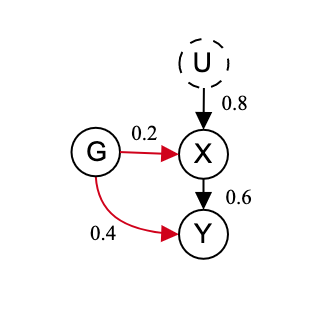}
\end{subfigure}
 \caption{Synthetic: Data generation graph for the synthetic data from ~\citet{yang2020causal} The nodes $U,X,Y$ are either follow a pareto or normal distribution. `U' denotes the utility or relevance of a datapoint with features `X` and `Y'. `G' denotes group information (e.g., male or female)  
 \label{fig:synthetic_dag} }
\end{figure}

The data generation process is as follows:  In each case, the causal model consists of the following nodes: node `X' and `Y' representing features or characteristics of a datapoint (item to be ranked in our case), `U' represents the relevance or utility, and `G' represents a protected group label. 
The generation process proceeds as follows: we sample from two relevance distributions (see  Fig.\ref{fig:synthetic_dag} in Appendix for all parameters) to create a set of candidates or subjects to rank. In each case, we simulate a dataset with 4 variables: a group variable `G' with values of `M' and `F' and each value has a probability of 0.5 to appear, i.e., 50\% is `M' and 50\% is `F'. The correlation among the two features `X' and `Y' are determined by `G' and `U' with pre-defined weights (here, `X' is determined by `G' and `U' with weights of 0.2 and 0.8 respectively;  `Y' is determined by `G' and `X' with weights of 0.4 and 0.6 respectively).
For values after binning, see Table~\ref{tab:ds_summary} in the main text. We used the \href{https://github.com/DataResponsibly/MirrorDataGenerator}{open-sourced implementation for user-specified causal relations} for all data generation.

\section{Reproducibility Details}
\label{app:reproducibility_details}
All simulations were carried out with the ULTRA-pytorch toolbox\footnote{\url{https://github.com/ULTR-Community/ULTRA_pytorch}}

\paragraph{Raw data} All data is stored in \emph{libsvm} format, and since we study global rankings, we consider a single-query setup. In the context of our datasets, this is analogous to querying with a specific string. We standardize all features, using median and interquartile range obtained from the training set in each case~\cite{pedregosa2011scikit}.

\paragraph{Learning algorithm} We used the IPS learning algorithm, with a neural network ranking model.   Under this model, we use pre-trained click models and propensity weights inferred using the randomization method proposed by ~\citet{wang2016learning} that randomly shuffles the top-k results and uses the average click-through rate at each
position to estimate the propensity. We use pre-trained propensity weights corresponding to the click model rather than re-training the pipeline twice.

\paragraph{Ranking model} In all cases, we use a neural network with 2 hidden layers consisting of 256 and 128 units each. A list-wise  loss is minimized with Stochastic Gradient Descent (SGD) optimizer and an initial learning rate of 0.01. Batch size is set to 256. The weights ranking model is checkpointed on the validation set every 10 iterations.  Finally, the checkpoint with highest NDCG@10 score on the validation set is used to generate rankings. Note that predicted softmax transformations of predicted logits are used to generate and sort items on the test set, and softmax activation is used with list-wise loss during training. Since softmax is a monotonic function, the use of logits for ranking--and hence, as ``relevance"-- is justified. However, results corresponding to criteria tests might change as a function of this, though we do not observe major changes in the settings we benchmark.

\paragraph{System evaluation} We evaluated all systems with multiple metrics such as NDCG on the test set averaged across 10 random seeds. In all metrics, the top-k positions were studied with $k=10$ unless specified otherwise. 
\paragraph{Fairness Metrics} We assume log-distribution dropping off at k=10 unless specified otherwise.

\section{Justification for Fairness Intervention Choice}
We use the relevance scores of items at group-level to compute the desired distribution of groups via the fair re-ranking algorithms proposed by ~\citet{geyik2019fairness}. While the unfairness mitigation algorithms don't consider the distribution of attention during re-ranking (and can only re-rank items such that the \emph{proportion} requirements of groups are met), ~\citet{geyik2019fairness} showed that these algorithms also improve the normalized discounted cumulative KL-divergence (NDKL) which computes a log-discounted KL-divergence between obtained and desired distribution of groups. We assume log-decaying attention distributions for measuring exposure, so we assessed this algorithm to match our setup well.

\end{document}